\def\year{2021}\relax
\documentclass[letterpaper]{article} 
\usepackage{aaai21}  
\usepackage{times}  
\usepackage{helvet} 
\usepackage{courier}  
\usepackage[hyphens]{url}  
\usepackage{graphicx} 
\usepackage{graphicx}
\usepackage{bm}

\usepackage{natbib}  
\usepackage{caption} 
\usepackage{booktabs}
\usepackage{dcolumn}
\usepackage{subcaption}
\usepackage{amssymb,amsmath}
\usepackage{cleveref}
\usepackage{xspace}
\usepackage{bbm}
\usepackage{tabularx}
\usepackage{booktabs}
\usepackage{mathptmx}
\usepackage{enumerate}    
\usepackage{paralist} 
\usepackage{fontawesome5}

\usepackage{cleveref}
\usepackage{xspace}
\usepackage{bbm}
\usepackage{tabularx}
\usepackage{booktabs}
\usepackage{paralist} 
\usepackage{fontawesome5}

\usepackage{marginnote}

\usepackage{xcolor}

\title{Stranger Danger!  Cross-Community Interactions with Fringe Users Increase the Growth of Fringe Communities on Reddit}

\author {
}

\title{Stranger Danger! Cross-Community Interactions with Fringe Users Increase the Growth of Fringe Communities on Reddit}
\author {
    Giuseppe Russo,\textsuperscript{\rm 1}
    Manoel Horta Ribeiro, \textsuperscript{\rm 2}
    Robert West \textsuperscript{\rm 2} \\
}
\affiliations {
    \textsuperscript{\rm 1} ETH Zurich \\
    \textsuperscript{\rm 2} EPFL \\
    giusepperusso@ethz.ch, manoel.hortaribeiro@epfl.ch, robert.west@epfl.ch
}
\makeatletter
\renewcommand*{\@textcolor}[3]{%
  \protect\leavevmode
  \begingroup
    \color#1{#2}#3%
  \endgroup
}
\makeatother

\usepackage{dcolumn,booktabs}
\newcolumntype{d}[1]{D{.}{.}{#1}}

\newcolumntype{C}{>{\centering\arraybackslash}X}

\usepackage[detect-all]{siunitx}

\usepackage{todonotes}

\begin{document}
\maketitle
\begin{abstract}
Fringe communities promoting conspiracy theories and extremist ideologies have thrived on mainstream platforms, raising questions about the mechanisms driving their growth. 
Here, we hypothesize and study a possible mechanism: new members may be recruited through \emph{fringe-interactions}: the exchange of comments between members and non-members of fringe communities. 
We apply text-based causal inference techniques to study the impact of fringe-interactions on the growth of three prominent fringe communities on Reddit: r/Incel, r/GenderCritical, and r/The\_Donald. 
Our results indicate that fringe-interactions attract new members to fringe communities. 
Users who receive these interactions are up to 4.2 percentage points (\textit{pp}) more likely to join fringe communities than similar, matched users who do not.
 This effect is influenced by  1) the characteristics of communities where the interaction happens (e.g., left vs. right-leaning communities) and 
2)  the language used in the interactions.
Interactions using toxic language have a 5\textit{pp} higher chance of attracting newcomers to fringe communities than non-toxic interactions. 
We find no effect when repeating this analysis by replacing fringe (r/Incel, r/GenderCritical, and r/The\_Donald) with non-fringe communities (r/climatechange, r/NBA, r/leagueoflegends), suggesting this growth mechanism is specific to fringe communities. 
Overall, our findings suggest that curtailing fringe-interactions may reduce the growth of fringe communities on mainstream platforms.

\end{abstract}

\section{Introduction}\label{sec:introduction}
Mainstream platforms enacted various moderation policies to curtail fringe communities due to their association with real-world violence and online harassment~\cite{qanon2020}.
For instance, r/The\_Donald, a Reddit community that was key in planning the 2021 US Capitol invasion \cite{trump_train}, was extensively sanctioned by Reddit, having its visibility reduced and being removed from the main feed.
However, even amidst sanctions, fringe communities still flourish and attract new members~\cite{trujillo2022make, horta2021platform}, e.g., even after the sanctions, r/The\_Donald remained one of the most active communities on Reddit, with over $790,000$ users before being banned. 
This observation suggests that visibility through platform affordances (e.g., Reddit's front page) may not be the sole mechanism driving the attraction of new members.

\begin{figure}[t]
    \centering
\includegraphics[width=\linewidth]{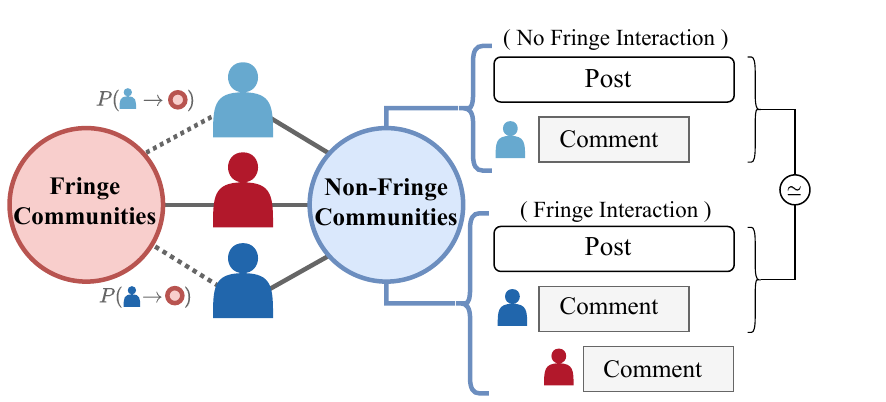}
    \caption{
    We study whether interacting with fringe users (\textcolor[HTML]{b2182b}{\faUser}) causes non-fringe users (\textcolor[HTML]{2166ac}{\faUser}) to join fringe communities. 
    We match users that involuntarily received comments from fringe users (\textcolor[HTML]{2166ac}{\faUser}; ``treatment'')  
    with users who did not, but had similar chances to do so (\textcolor[HTML]{67a9cf}{\faUser}; ``control;''  \textcolor[HTML]{67a9cf}{\faUser} $\simeq $ \textcolor[HTML]{2166ac}{\faUser}). 
    Then, we estimate the causal effect of fringe interactions by comparing the fraction of users that went on to join the fringe community in the treatment ($P($  \textcolor[HTML]{2166ac}{\faUser} $\rightarrow$ \textcolor{red}{$\bullet$}$)$) 
    and in the control group ($P($\textcolor[HTML]{67a9cf}{\faUser} $\rightarrow$ \textcolor{red}{$\bullet$}$)$).
    }
    \label{fig:interaction}
\end{figure}

\vspace{1.5mm}
\noindent
\textbf{Present work.}
Here, we consider another potential mechanism for the growth of fringe communities on mainstream platforms: \emph{fringe-interactions}, defined as an exchange of comments between users of fringe communities and other users active on the same platform (e.g., Reddit) but not in the fringe community (e.g., r/The\_Donald) at the time of the interaction.
We hypothesize that fringe-interactions boost the visibility of fringe communities by exposing non-fringe users to ideas or concepts associated with the fringe community via either the profile of fringe users or the content of the interactions.

We center our analysis around three research questions:
\begin{itemize}
    \item[] \textbf{RQ1}: Do fringe-interactions drive newcomers toward fringe communities? If so, is this specific to fringe communities?
    \item[] \textbf{RQ2}: In which communities are fringe-interactions more successful in attracting newcomers?
    \item[] \textbf{RQ3}: What linguistic traits characterize fringe-interactions that successfully attract newcomers?
\end{itemize}

To answer these questions, we study three prominent fringe communities on Reddit (r/Incel, r/GenderCritical, and r/The\_Donald) using the quasi-experimental setup illustrated in \cref{fig:interaction}.

First (\textbf{RQ1}), we use regression analysis to measure the probability of users joining fringe subreddits after interacting with fringe users. 
We then compare this probability to that of users who did \emph{not} interact with fringe users. 
Via balanced risk set matching \cite{rosenbaum1983central}, we ensure that these two groups of users have comparable propensities to interact with fringe users.
To understand if this mechanism is specific to fringe communities, we repeat the analysis, considering interactions between members of \emph{non-fringe} subreddits (r/climatechange, r/NBA, r/leagueoflegends) and other users not yet active in these subreddits.
Second (\textbf{RQ2}), we analyze how the effect of fringe-interactions varies depending on the characteristics of the community (e.g., political leaning) where the fringe-interaction occurred. 
Third (\textbf{RQ3}), we characterize the content of fringe-interactions that successfully attract newcomers across a variety of linguistic traits (e.g., toxicity).

\vspace{1mm}
\noindent
\textbf{Findings.} 
Our analysis shows that fringe-interactions increase the likelihood of a user posting on a fringe community by up to 4.2 percentage points (\textit{pp}) (\textbf{RQ1}). 
We do not find evidence that interactions with non-fringe users (e.g., members of r/climatechange) drive newcomers towards non-fringe communities (r/climatechange itself), suggesting that the growth through interactions is specific to fringe communities.

Additionally, we find that the strength of the effect of fringe-interactions varies depending on the characteristics (political orientation, gender, and age) of the subreddits where the interactions occur  (\textbf{RQ2}). 
For instance, users interacting with fringe users in right-leaning subreddits are 15.6 \textit{pp} more likely to join fringe communities than those active in the same subreddit who have not interacted with fringe users.
Finally, our analysis indicates that interactions containing toxic language increase the chances of attracting newcomers to fringe communities by 5\textit{pp} on average (\textbf{RQ3}). 

\vspace{1mm}
\noindent
\textbf{Implications.}
Mainstream online platforms have attempted to curtail fringe communities by banning them or reducing their visibility.
However, neither of these interventions is a silver bullet, as they are remarkably resilient~\cite{horta2021platform, trujillo2022make}.
The growth mechanism studied here shows that fringe-interactions are another target for stakeholders. 
For instance, platforms could reduce the visibility of fringe community members \textit{outside} of fringe communities or proactively target users who receive messages from fringe community members with information that may reduce their chance to join fringe communities.

\begin{table*}[ht]
\caption{Data collection summary --- For the three fringe subreddits considered in this paper, we show the number of comments, users, and users obtained (columns 2--4), as well as the treated and potential control units associated with these users' fringe-interactions (columns 6--7) across one-year periods (column 5). We also show the same statistics for non-fringe subreddits that we use to repeat the experiment.}
\label{tab:data_summary}

\centering
\begin{tabular}{lrrrrrr}
\toprule

 & Comments & Users & Selected Users &  Period Considered & 
 Treated & Potential Controls \\
\midrule
\emph{Fringe subreddits} &  &  &   &   &  & \\ \midrule

r/Incels & $2{,}041{,}313$ & $177{,}095$ & $38{,}145$ & $06/2016-06/2017$ & $86{,}556$ & $5{,}275{,}321$\\
r/GenderCritical & $1{,}629{,}169$ & $48{,}243$ & $9{,}624$ & $01/2019-01/2020$ & $54{,}142$ &  $2{,}250{,}641$\\
r/The\_Donald & $12{,}387{,}349$ & $482{,}796$ & $72{,}371$ & $05/2018-05/2019$ & $97{,}823$ & $6{,}481{,}223$\\
\midrule
\emph{Non-Fringe subreddits} & & & & & &\\\midrule
r/climatechange & $632{,}674$ & $93{,}256$ & $19{,}374$ & $01/2019-01/2020$ & $81{,}307$ & $3{,}912{,}866$\\
r/NBA & $1{,}483{,}631$ & $468{,}321$ & $67{,}481$ & $01/2018-01/2019$ &  $72{,}321$ & $2{,}973{,}364$\\
r/leagueoflegends & $3{,}573{,}974$ & $694{,}522$ & $71{,}944$ & $01/2019-01/2020$ &  $95{,}763$ & $6{,}514{,}447$\\
\bottomrule
\end{tabular}
\end{table*}

\section{Related Work}
\label{sec:rel_work}

\vspace{.5mm}
\noindent
\textbf{Why do people join online communities?} 
Scholars have extensively studied user motivations for joining online communities~\cite{ridings2004virtual, ren2012building}.
Past research indicates that users deliberately seek communities that align with their social identities~\cite{ammari2015understanding, lingel2014city}, and has highlighted the importance of content quality~\cite{lu2011encouraging,  zhang2017community, russo2023acti},  effective moderation~\cite{lampe2005follow}, and meta-characteristics like size, activity levels, and network structures \cite{hwang2021people}.
Further, \citet{backstrom2008preferential} has found that interpersonal interactions with other users draw individuals to online communities.
In the case of fringe communities, research on why users participate in these communities focused on users' psychological characteristics \cite{schmid2013radicalisation}, finding that hopelessness, sadness, and anxiety are the primary psychological drivers of participation~\cite{caren2012social, 10.1145/2470654.2470702}.

\vspace{.5mm}
\noindent
\textbf{Fringe communities.}
We briefly describe the fringe communities considered in this study.
r/The\_Donald,  created in June 2015 to support Donald Trump's presidential campaign, became associated with the "alt-right" movement, hosting discussions involving racism, sexism, and Islamophobia. It also spread conspiracy theories and was mobilized for ``political trolling'' \cite{lyons2017ctrl, paudel2021soros}.
r/GenderCritical, created in September 2013, hosted the trans-exclusionary radical feminists (TERFs) community, known for doxing and harassing trans women \cite{atlantic,williams2020ontological}.
r/Incel, created in August 2013, was a community of self-denominated "involuntary celibates" adhering to ``The Black Pill,'' the belief that unattractive men are doomed to romantic loneliness and unhappiness \cite{ribeiro2021evolution}. 
Since their inception, the Incels community has been closely related to terrorist attacks and the production of misogynistic content online \cite{jaki2019online, hoffman2020assessing}.

\vspace{.5mm}
\noindent
\textbf{Interventions against fringe communities.} 
 r/Incel, r/The\_Donald, and r/GenderCritical have all been banned due to breaching Reddit's guidelines.
 Previous work has studied the effectiveness of these bans, showing that banned users reduce their activity on mainstream platforms~\cite{chandrasekharan2017you,jhaver2021evaluating}. However, 
part of the banned community migrates to other fringe platforms~\cite{monti2023online, russo2023understanding} causing possible spillovers of antisocial behavior back onto the mainstream platform \cite{russo2023spillover, schmitz2022quantifying}. 
 Other works have examined softer interventions like limiting the visibility of fringe communities and warning new visitors about the potential issues with content posted in these communities~\cite{chandrasekharan2022quarantined,trujillo2022make}, finding that these measures reduce the number of newcomers to some extent.  
 Overall, this literature suggests that interventions against fringe communities are no silver bullet, as they are highly adversarial and have highly engaged members. 
 
\vspace{.5mm}
\noindent
\textbf{Relation between present and prior work.}
Here, we explore fringe-interactions, a potential mechanism that may contribute to the growth of fringe communities.
The existence of this mechanism is backed by previous research highlighting the importance of interpersonal interaction in drawing users to new online communities~\cite{backstrom2008preferential}.
Given how challenging it is to reduce the influence of fringe communities in our online ecosystem~\cite{horta2021platform, monti2023online}, we argue that understanding why these communities thrive is key to increasing the arsenal of interventions available to curtail their growth.

\section{Material and Methods} \label{sec:methods}

\subsection{Data}

\vspace{1.5mm}
\noindent
\textbf{Reddit data.}
To answer our research questions, we selected three prominent fringe subreddits r/Incels,  r/GenderCritical, and r/The\_Donald (see \Cref{sec:rel_work}).
Using the Pushshift API~\cite{baumgartner2020pushshift}, we retrieve all comments made in these three fringe subreddits from their creation to their banning from Reddit.
We consider as part of r/Incels,  r/GenderCritical, or  r/The\_Donald only those users that post more than five comments in one of these three subreddits. 
This is similar to what has been done in previous research~\cite{kumar2018mega, samory2018conspiracies}.
In cases where a user exceeds the threshold in multiple fringe subreddits, we consider the user as a member of the subreddit where they have posted the most; this prevents fringe users from being considered multiple times across different fringe subreddits.
We collect roughly 15M posts from 700K users from these three fringe subreddits. 

To understand if the growth through interactions is specific to fringe communities, we repeat our analyses, replacing fringe with \emph{non}-fringe subreddits (r/climatechange, r/NBA, and r/leagueoflegends). 
Therefore, we collect all posts made on three non-fringe subreddits, r/climatechange, r/NBA, and r/leagueoflegends, chosen to represent a diverse set of communities.
\footnote{Respectively, a political movement, a community centered around an ``offline'' event, and a community centered around an online video game.}
We collect roughly 5M posts from 1M users from these three non-fringe subreddits. 
We provide statistics for these collection processes in \Cref{tab:data_summary}.

\vspace{1.5mm}
\noindent
\textbf{Treatment and control groups.}
Our study centers on understanding how interactions with fringe users influence the attraction of newcomers to fringe subreddits.
To achieve this, we compare the likelihood of users joining a fringe subreddit after a fringe-interaction (\emph{treatment group}) with the likelihood of other users joining the same fringe subreddit without any interaction (\textit{control group}).
To identify users in the treatment group, we gather all fringe-interactions made by users of fringe subreddits on subreddits other than fringe ones. 
In other words, we look for interactions occurring in subreddits different from r/Incels,  r/GenderCritical, or r/The\_Donald or related to them (e.g., Incels2).

A fringe-interaction consists of a comment $c_{\text{fringe}}$ made by a fringe user $u_{\text{fringe}}$ on a non-fringe subreddit  $s_{\text{non-fringe}}$ in response to a comment $c_{\text{non-fringe}}$ made by a non-fringe user $u_{\text{non-fringe}}$ on the same non-fringe subreddit $s_\text{non-fringe}$.
We include non-fringe users in our treatment group if the following conditions are met:

\vspace{1mm}
\begin{compactenum}
    \item the non-fringe user $u_{\text{non-fringe}}$ did not post in the fringe subreddit (e.g., r/Incels) associated with the fringe user $u_{\text{fringe}}$ prior to the interaction. 
    \item $u_{\text{fringe}}$ has never interacted before with $u_{\text{non-fringe}}$. In other words, they never exchanged a comment
    \item The comment $c_{\text{fringe}}$, authored by $u_{\text{fringe}}$, occurred within a week from the comment $c_{\text{non-fringe}}$ posted by $u_{\text{non-fringe}}$.
\end{compactenum}
\vspace{1mm}

Finally, following previous research from \citet{phadke2022pathways}, we collect our data only from subreddits that received at least five contributions from users of fringe subreddits and consider fringe-interactions that happened within one year.
To define a control group, we collect all comments made in the same weeks and subreddits as the ones in the treatment group where no fringe-interaction happened.
We provide these statics in \Cref{tab:data_summary}

\vspace{1.5mm}
\noindent
\textbf{Outcome.}
Given users that receive (treatment) or not (control) an interaction from a user active on a fringe subreddit (e.g., r/Incels), our outcome variable is the number of users joining the same fringe subreddit in the weeks following the interaction. 
Specifically, we analyze users' post and comment activity in the eight weeks following the date of comments associated with users in our treatment/control groups.
We consider that a user ``joined'' a fringe subreddit if they post or comment at least once in that subreddit after the interaction.
We operationalize these covariates following the same approach when repeating our analysis considering the interactions of non-fringe users from r/climatechange, r/NBA, and r/leagueoflegends.

\subsection{Identification}

Our analysis uses observational data to mimic a hypothetical experiment.
Suppose Reddit implemented a filter where whenever a user who is active in a fringe community replies to someone in a non-fringe community, the comments of the fringe user do not get shown.
Then, for a select, randomly assigned treatment group, this filter gets removed.
Conversely, nothing changes for the remaining control users; the filter remains as is.
We are interested in the difference in the rate at which treatment vs. control group users join the fringe community.

\begin{figure}[t]
    \centering
\includegraphics[scale=1]{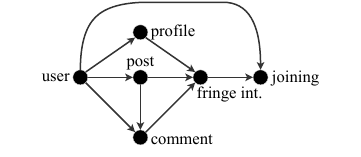}
    \caption{Causal diagram. Given a comment $c_{\text{non-fringe}}$ made by a non-fringe user $u_{non-fringe}$, we assume that a fringe-interaction depends exclusively on (1) the \textit{profile} of the non-fringe who posted the comment $c_{\text{non-fringe}}$, (2) on the \textit{post} where the comment was made, and (3) on the content of the \textit{comment} itself.
    We control for these factors to estimate the effect of a fringe-interaction in subsequent participation in a fringe community (\textit{fringe\ int.}\ $\rightarrow$ \textit{joining}).
    }
    \label{fig:causal_dag}
\end{figure}

Since this experiment is not feasible, we use observational data as a substitute. 
Our treatment group consists of users who interacted with a fringe user from either r/Incels, r/GenderCritical, or r/The\_Donald.
On the other hand, our control group consists of users who were equally likely to interact with a fringe user but did not.
To identify the effect of fringe-interactions on non-fringe users, we formalize our assumptions in the causal graph in \cref{fig:causal_dag}. 
We assume that given a comment $c_{\text{non-fringe}}$ (\textit{comment} in the causal graph) whether it receives a fringe-interaction depends on (1) the profile of the user $u_{\text{non-fringe}}$ (\textit{profile} in the causal graph) who posted the comment $c_{\text{non-fringe}}$ ,
(2) the content of the comment $c_{\text{non-fringe}}$ (\textit{comment}), and
(3) the content  of the post $p$ (\textit{post}) that corresponds to the  Reddit submission the comment $c_{\text{non-fringe}}$ refers to.

Under these assumptions, we can control for confounders by blocking the biasing paths ~\cite{glymour2016causal} to isolate the causal effect of fringe-interactions.
We do so by controlling the user profiles, the content, and the structure of comments and posts, respectively.
Indeed, different comments/posts may draw attention differently, stimulating possible interactions from fringe users. 
However, other confounders may affect the likelihood of non-fringe users to join fringe subreddits \textit{regardless of fringe-interactions}, e.g., being a young male with relationship issues makes someone more likely to join r/Incels, we address this issue in detail in \Cref{sec:robustness}.

\subsection{Operationalization}\label{sec:methods:covariates}
Below, we describe how we operationalize the confounding variables in the causal diagram in \Cref{fig:causal_dag}. Note that these covariates are computed for a triplet $\langle \text{comment},\text{post},\text{profile}\rangle$, where, in the treatment group, the user who made the comment interacted with a fringe user, and in the control group, they did not.

\vspace{3mm}
\noindent
\textbf{Comment.} \label{sec:methods:covariates:content}
When comparing comments that received fringe-interactions with those that did not, we want the content of these comments to be semantically similar.
In a recent comparison of text-level adjustment strategies, \citet{weld2022adjusting} found that transformer-based representations outperformed other text representations, and thus, we use a fine-tuned version of the BERT architecture \cite{devlin2018bert} to obtain a representation for each comment in our dataset. We describe this in detail in \Cref{sec:methods:effect:brsm}.

\vspace{3mm}
\noindent
\textbf{Post.} \label{sec:methods:covariates:post}
 We also want the posts associated with comments (potentially) receiving fringe-interaction to be similar.
This is partially addressed by matching comments within similar subreddits, but in addition, we also consider, for each post, the number of (1) direct replies that a post received, (2) unique users that directly replied to the post, (3) comments in the post, and (4) unique user commenting in the post. 
Importantly, these variables are computed prior to the creation of the comment of interest.

\vspace{3mm}
\noindent
\textbf{Profile.} \label{sec:methods:covariates:user}
Fringe users may choose to interact with a non-fringe user (via a comment) because of the profile of the non-fringe user that posted the comment.
To control for the profile, we characterize each non-fringe user using their activity and its relatedness to the fringe subreddit (e.g., r/Incels).
Specifically, we operationalize (1) \textit{user activity} as the total number of posts made in the $8$ weeks before the comment, and (2) the \textit{fringe score} as the proportion of the user’s Reddit activity dedicated to discussion related to the fringe subreddit. To calculate the \emph{fringe score}, we follow \citet{phadke2022pathways}: for a non-fringe user $u$ that write $N_u$ posts in a set $S_u$ of subreddits the fringe score $f_u$ is 
 \begin{equation}
  f_{u}=\frac{\sum_{s \in S_u}n_{s}\text{sim}(s_\text{fringe},s)}{N_u},
\end{equation}
where $n_{s}$ is the number of comments made on the subreddit $s$, $\text{sim}(s_{\text{fringe}},s)$ is the cosine similarity between the embeddings (from \citet{waller2021quantifying}) of the fringe subreddit $s_{\text{fringe}}$, (e.g., r/Incels) and a subreddit $s$.

\subsection{Estimation}\label{sec:methods:effect}

Having operationalized the treatment, the outcome, and the control variables of interest, we then estimate the causal effect of fringe-interactions. Our approach is two-fold: we match comment-user pairs ($c_{non-fringe}$, $u_{non-fringe}$)  that received fringe-interactions with similar pairs that did not.
We then conduct a regression analysis in the subset of matched comments posted by non-fringe users.
This approach allows us to obtain robust results by combining the strengths of both approaches~\cite{matchingereg}.

\begin{figure}[t]
    \centering
\includegraphics[width=0.95\columnwidth]{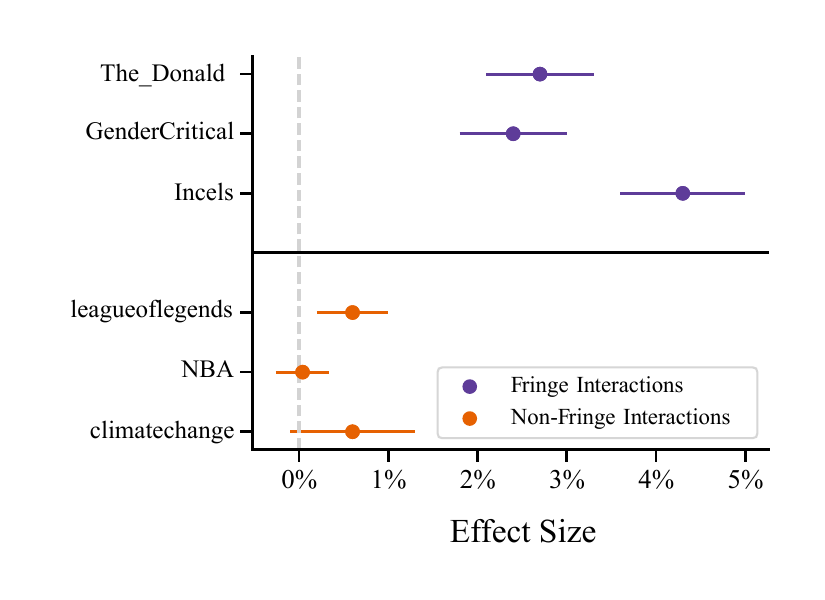}
    \caption{The effect of fringe-interactions on joining fringe communities--- Differential increase in the probability of joining fringe subreddits after interacting with fringe users. Significant effect observed for r/Incels, r/GenderCritical, and r/The\_Donald.
    We repeat the experiment considering interactions with users of non-fringe communities finding not significant or very small effects. Significance level at $0.05$}
    \label{fig:total_effect}
\end{figure}

\begin{figure*}[t]
    \centering
\includegraphics[width=1.0\textwidth]{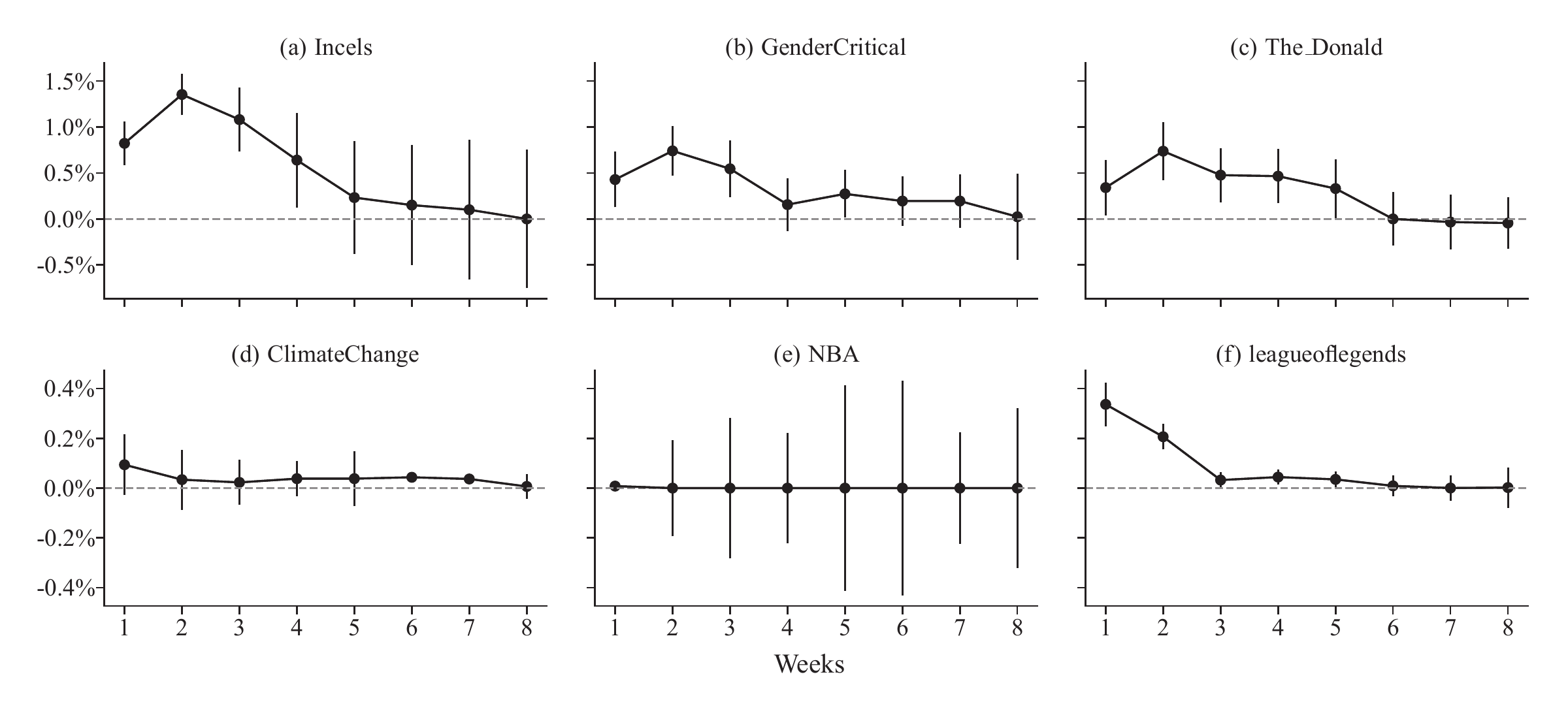}
    \caption{Top row shows the effect size of fringe-interactions on users that interacted with  members of fringe communities (r/Incels, r/GenderCritical, r/The\_Donald). Bottom row shows the effect of interactions when users interact with members of non-fringe communities (r/climatechange, r/NBA, r/leagueoflegends). The effect is stratified over a follow-up period of eight weeks ($x$-axis). Note that the y-axis scale is different between top and bottom row. }
    \label{fig:rq1_temporal}
\end{figure*}
\vspace{-0.1cm}

\vspace{1.5mm}
\noindent
\textbf{Balanced Risk Set Matching.}\label{sec:methods:effect:brsm}
In our setting, users in the treatment group receive the treatment (fringe-interactions) at different times. 
 This is because fringe users reply to comments of non-fringe users at different times.
When a treatment is given at various times, it is important to form matched pairs in which subjects are similar prior to treatment but avoid matching on events subsequent to treatment. 
This is done using risk-set matching, in which a newly treated individual at time $t$ is matched to one control not yet treated at time $t$ based on covariate information describing subjects prior to time $t$.
The covariates we must control for with the matching (see \Cref{fig:causal_dag}) vary with time, e.g., a non-fringe user  have a similar profile to another non-fringe user only in time $t_1$, not $t_2$.

We address this issue by performing Balanced Risk Set Matching~\cite{rosenbaum1983central}, a procedure that consists of 
(1) creating ``risk sets,'' groups of individuals that did not receive the treatment up to a point in time; 
(2) estimating the propensity score of individuals in the risk sets, i.e., the probability of being treated given pre-treatment characteristics; and 
(3) using the propensity score to match individuals who went on to receive the treatment (here, fringe-interactions) with those who did not.
We describe this process below.

\begin{compactenum}
    \item \textit{Creating risk sets --} 
    Each treated user $u_{non-fringe}$ is treated at a different treatment time (i.e., receive a fringe-interactions), where the treatment time corresponds to the day a treated user $u_{non-fringe}$ posted on the subreddit $s_{non-fringe}$ the comment $c_{non-fringe}$  that received the fringe-interaction.
    We subsample the set of possible control users to those controls that posted a comment $c'_{non-fringe}$ on the same subreddit $s_{non-fringe}$ on the same week of the interaction (treatment).
    By following this procedure, we create comparable sets of users.
    We represent each of these users considering the comment-, post-, and profile-related covariates calculated eight weeks before the treatment time $t$ of each risk set.
    
    \item  \textit{Training propensity score model --} We compute the propensity to receive an interaction from a fringe user $u_{fringe}$ for each comment $c_{non-fringe}$ written by a treated or control user $u_{non-fringe}$ in a risk set.
    Specifically,  we fine-tune BERT \cite{devlin2018bert} with an additional linear layer to compute the propensity that a comment written by a user receives a fringe-interaction, providing as input to the model the text of the content $c_{non-fringe}$ and the title of the corresponding Reddit submission.
 Before the final linear layer, we concatenate the [CLS] token with user and post-level covariates (computed considering the user $u_{non-fringe}$ activities in the eight weeks before time $t$) to compute the propensity of a comment $c_{non-fringe}$ to receive a comment from a fringe user $u_{non-fringe}$. 

    \item \textit{Matching --} Finally, we match each treated user to a control user (within in each risk set) using the nearest neighbor algorithm on the propensity score considering a calliper of $0.001$. Further, for each treated user, we only consider control users who posted in the same subreddit in the same week.
    This procedure yields $82{,}725$ pairs of users, and there were $3,831$ treated users we could not match.
    We assess the quality of the matching by measuring the standardized mean difference of user-, comment- and post-level covariates, obtaining absolute standardized mean differences smaller than the standard 0.1 threshold~\cite{austin2011introduction} for most covariates considered (13/15 for r/Incels, 14/15 for r/GenderCritical, and 12/15 for r/The\_Donald (see ~\Cref{app:robustness} ). 

\end{compactenum}
\noindent
We provide examples of matched comments written by the treated and control users that we match according to our matching procedure in \Cref{tab:examples}.

\vspace{1.5mm}
\noindent
\textbf{Regression Analysis.}
Considering matched users, we estimate the effect of fringe-interactions with the linear model

\begin{equation}
    y_{ut} = \beta_0 + \beta_1\text{treated}_{ut} + \pmb{\gamma}^{T}\textbf{X}_u + F_t,
    \label{eq:basic_regression}
\end{equation}

where $y_{ut}$ represents a binary variable indicating whether a \emph{non-fringe} user $u$ joined a fringe community in the eight weeks after the fringe-interaction at week $t$;
$\text{treated}_{ut}$ indicates if user $u$ received the treatment (fringe-interaction) in week $t$. 
Therefore, $\beta_1$ captures the effect of the fringe-interaction on the non-fringe user $u$. 
$\textbf{X}_u$ represents the array of control variables introduced in \Cref{sec:methods:covariates} at the user level computed in the pre-treatment period; and 
$F_w$ are weekly fixed effects, which control for potential latent time trends that could impact our results. 
Note that, coefficients here must be interpreted as percentage point (\textit{pp}) increases.

\section{Results}\label{sec:results}

\subsection{RQ1: Effect of fringe-interactions}

\vspace{1.5mm}
\noindent
\textbf{Overall Interaction Effect.} 
Non-fringe users who receive a fringe-interaction exhibit a significant increase in the probability of joining the said fringe community relative to those who do not (see \Cref{fig:total_effect}).
Specifically, considering the regression analysis described in \Cref{eq:basic_regression}, we find that fringe-interactions with r/Incels users increase by 4.2 $pp$ the likelihood of posting on r/Incels in the eight weeks following the interaction with a fringe user of r/Incels (2.4 $pp$ for r/GenderCritical; 2.2 $pp$ for r/The\_Donald).
To understand if such an effect is specific to fringe communities, we repeat the same analysis considering interactions with users of non-fringe communities.
Practically, we substitute the three fringe subreddits with r/climatechange, r/NBA, and r/leagueoflegends. We then repeat our observational study.
We find effects to be smaller and not statistically significant,  0.5 $pp$ for r/ClimateChange and r/leagueoflegends; and 0.0006 $pp$ for r/NBA.

\vspace{1.5mm}
\noindent
\textbf{Time-Stratified Interaction Effect.}
To study the variation of the effect within the observation window, we re-run \Cref{eq:basic_regression} considering the data associated with each week in the post-intervention follow-up period, i.e., we estimate the effect of the user joining the community on the first, second, third week and so on after the fringe-interaction.
For the fringe communities considered (\Cref{fig:rq1_temporal}; top row), we find that the effect is strongest during the second week following the interaction with fringe users (1.3 $pp$ for r/Incels; 0.7 $pp$ for r/GenderCritical; 0.8 $pp$ for r/TheDonald; all results statistically significant with $p<0.05$). 
This likely happens because the very first week can encompass fewer days in our setup (e.g., if the external interaction happened on a Friday, fewer days are being considered in week \#1).
After the second week, the effect gradually wanes for all three communities.
For the non-fringe communities considered (\Cref{fig:rq1_temporal}; bottom row), weekly effects are only significant for the r/leagueoflegend community, which, interestingly, does not exhibit the peak in the second week.

\begin{figure*}[t]
    \centering
\includegraphics[width=1.0\textwidth]{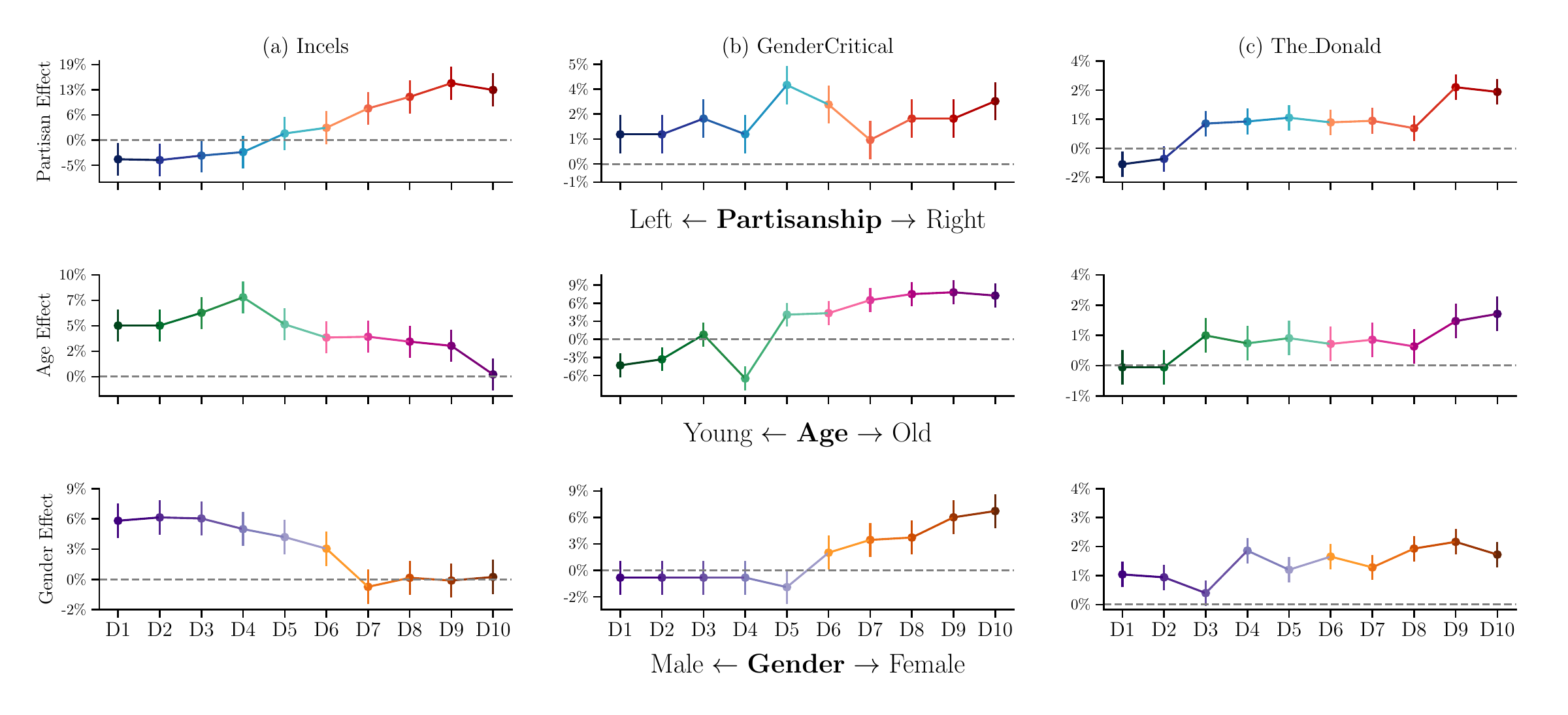}
    \caption{\emph{Stratification of External Interaction Effects Based on Social Dimension Scores.} Effects of fringe-interactions are categorized according to the partisan  (top row), age (middle row), and gender  (bottom row) scores of the subreddits where the fringe-interaction occurred. Subreddits are grouped into ten distinct categories (D1-D10), see \Cref{tab:social_scores}, according to their social dimension scores ($x$-axis).  Each column illustrates the effect of fringe-interactions with  users of  r/Incels, r/GenderCritical, or r/The\_Donald. All results statistically significant with $p<0.05$
    }
    \label{fig:stratification}
\end{figure*}

\subsection{RQ2: In which subreddits are fringe-interactions successful?}\label{sec:rq3}
To investigate which characteristics make subreddits more susceptible to the effect of fringe-interactions , we categorize subreddits across three ``social dimension scores'' as computed by \citet{waller2021quantifying}.
These represent the social positioning based on the partisanship, age, and gender of a subreddit.
Each score is a value between $-1$ and $+1$, describing how left or right, old or young, and how feminine or masculine a subreddit is.

We assign subreddits where a fringe-interaction occurred in one of ten groups (D1-D10). 
These groups are obtained by (1) computing the ten deciles of the social scores and (2) assigning each subreddit to one of these groups based on its social score. We repeat this procedure for all three social dimensions. 
To study in which subreddits fringe-interactions are most effective, we run the following regression:

\begin{equation}
    y_{u}=\beta_0 + \beta_1 \text{treated}_u + \pmb{\beta}^{T}  \pmb{D}_{s(u)} \, \text{treated}_u + \pmb{\gamma}^{T}\textbf{X}_u + F_t
     \label{eq:strat_regression}
\end{equation}

here $D_{s(u)}$ is a one hot encoded vector that is zero everywhere but in the position of the decile assigned to the subreddit $s$ where the non-fringe user $u$ received the fringe-interaction.
We index this subreddit with $s(u)$. 
Therefore, the vector of coefficients $\pmb{\beta}$ captures the effect of an interaction occurred in a subreddit associated in a specific group (D1-D10). 
The other variables are as described in \Cref{eq:basic_regression}. 


Considering the partisan dimension, interactions with r/Incels and r/The\_Donald users are most successful in right-leaning subreddits (15.3 $pp$ D9 for r/Incels) and (2.5 $pp$ D9 for r/The\_Donald interactions). For example, interactions are effective in subreddits like r/GunsForSale, r/russia, and r/totalwar which are part of D9 group.
Differently, fringe-interactions with r/Incels and r/The\_Donald users are least successful in left-leaning subreddits, where we found statistically significant negative effects.
Examples of subreddits in the D1 and D2 along the partisan scores are r/democrats, r/Marijuana, and r/Impeach\_Trump.
No similar political divide is observed for r/GenderCritical.

For the age dimension, interactions with r/Incels users are most successful on ``young'' subreddits (D1-D4) with an average effect of 6.2 $pp$, whereas for interactions with r/GenderCritical and r/The\_Donald users, the effect is more pronounced in ``older'' subreddits (D5-D10) with average scores of 6.6 $pp$ and 1.5 $pp$, respectively. 
Interestingly, the effect on young subreddits is negative for r/GenderCritical ($-3.3$ $pp$)  and negligible for r/The\_Donald  (0.06 $pp$) .

For the gender dimension, interactions with r/Incels users show higher effects in masculine subreddits, while r/GenderCritical interactions have a higher impact on feminine subreddits. No clear trend is observed for interactions with r/The\_Donald users.

\subsection{RQ3: The Effect of Linguistic Traits}

Last, we investigate whether linguistic traits of fringe-interactions (i.e., comments from fringe to non-fringe users) impact their effectiveness in attracting newcomers with the following linear model:
\begin{equation}
    y_u=\beta_0 + \beta_1 \text{treated}_u + \pmb{{\beta}}^{T}  \pmb{L_{c(u)}}  \,  \text{treated}_u + \pmb{\gamma}^{T}\textbf{X}_u + F_t
\end{equation}

where $L_{c(u)}$ represents the vector of linguistic traits of the comment $c$ written by the fringe user and received by the non-fringe user $u$ in the fringe-interaction (we index this comment with $c(u)$). The vector of coefficients $\pmb{{\beta}}$ represents the effect of each linguistic trait considered.
The other variables are as described in \Cref{eq:basic_regression} and \Cref{eq:strat_regression}.

Specifically, the linguistic traits we consider are 
(1)  anxiety, sadness, and ``they vs.\ we'' language computed accordingly to LIWC \cite{pennebaker2001linguistic},
(2) toxicity as computed by Perspective's API \cite{perspective} and binarized by considering texts with a score above 0.8 as toxic
(similarly to  \citet{ribeiro2021evolution}),  and 
(3) community-specific lingo (see \Cref{app:lingo}). 
We chose these linguistic traits because involvement in fringe groups is linked to an underlying psychological disposition that includes tendencies to experience emotions such as sadness and anxiety, as well as receptiveness to fringe narratives (e.g., community lingo and highly toxic content) \cite{chandrasekharan2017you, butter2020routledge}.
We code all these traits to assign values of $1$ if the linguistic trait is present in the interaction and $0$ otherwise.

\begin{figure}[t]
    \centering
    \includegraphics[width=\columnwidth]{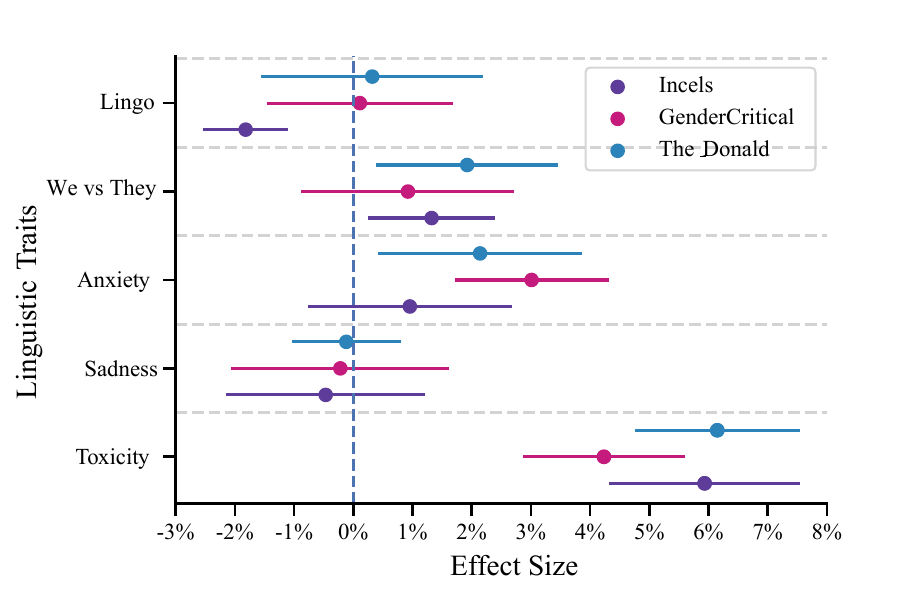}
    \caption{The effectiveness of linguistic traits of fringe-interactions --- On the y-axis, the linguistic traits considered. On the x-axis, the effect size of each linguistic traits associated with each fringe subreddit. }
    \label{fig:language}
\end{figure}

\Cref{fig:language} shows the observed effects for selected linguistic traits.
Toxic fringe-interactions increase the attraction of newcomers by $5.8$, $4.2 , 6.1$ $pp$ for r/Incels, r/GenderCritical, and r/The\_Donald, respectively.
Also, we find that comments containing anxious messages and ``they vs.\ we'' have a positive effect on attracting newcomers. 
Interestingly, for r/Incels,  we find a negative effect for fringe-interactions containing community-specific lingo. 

\section{Robustness Checks}\label{sec:robustness}

\subsection{Content-Level Exact Matching}\label{sec:robustness:exact}
Our findings rely on the semantic similarity of comments we matched using the approach described in \cref{sec:methods:effect:brsm}. 
To ensure the validity of our results, we established a strict criterion that the content of treated and control interactions (i.e., comments) must be exactly the same.
To fulfil this requirement, we collected 2,312 treated (received a fringe-interaction)  and 5,627 control comments with the same hyperlink as their sole content. 
Due to possible user and post-level confounders, we matched comments that received an interaction with those that did not, taking into account all user and post-level covariates.
We obtain $2,274$ pairs of matched comments considering fringe-interactions from r/Incels, r/GenderCritical, and r/The\_Donald together.
We run our regression analysis, finding average treatment effect of 1.3 $pp$ with ($p < 0.001$).

\subsection{Sensitivity Analysis}\label{sec:robustness:sensitivity}
Our results rely on the assumption that the treatment assignment is not biased. 
Meaning that the only difference between treated and control users is a simple coin flip.
Sensitivity analysis is a way of quantifying how the results of our study would change if this assumption is violated \cite{rosenbaum2005sensitivity},
This notion is quantified by the sensitivity $\Gamma$, which specifies the ratio by which the treatment of two matched persons would need to differ to result in a $p$-value above the significance threshold. Large values of $\Gamma$ corresponding to more robust conclusions. 
For the chosen $p=0.05$, we measured for the three fringe communities that we study $\Gamma$s of $2.3$, $1.35$, and $1.65$, which implies that, within matched pairs, an individual's probability of being the treated one could take on any value between $1/(1+\Gamma)=0.3$ and $\Gamma/(1+\Gamma)=0.69$ for r/Incels, between 0.42 and 0.57 for r/GenderCritical, and between 0.38 and 0.62 for r/The\_Donald  without changing our decision of rejecting the null hypothesis of no effect. 

\section{Discussion}\label{sec:discussion}

Understanding how fringe communities grow holds immense significance for mainstream platforms and policy-makers alike -- it is a stepping stone for interventions aimed at limiting their (well-documented) harm.
While previous work has pointed at algorithmic visibility and platform affordances as means by which fringe communities grow~\cite{munger2022right}, here we hypothesize and study another, more social mechanism.
We show that users who involuntarily receive interactions from fringe users increase their likelihood of participating in their fringe communities, and that this effect is modulated by where the interaction happens and what is said by the fringe user.

\vspace{1.5mm}
\noindent
\textbf{Implications.} Our results raise an important question: Is this mechanism relevant enough to warrant the attention of moderation policies? Within the observation period considered, $40,321$, $32,306$, and $123,562$ users joined r/Incels, r/GenderCritical, r/The\_Donald, respectively.
Based on our estimated effects, approximately 7.2\%, 3.1\%, and 2.3\% of the newcomers joined after interacting with users from r/Incels, r/GenderCritical, or r/The\_Donald.
This observation suggests that community-level moderation policies could be combined with sanctions applied to individual users. 
These sanctions might include reducing the visibility of their posts or limiting the number of comments they can make in more susceptible communities. 
Such a combination could diminish the impact of fringe-interactions and slow down the growth of fringe communities on mainstream platforms.

\vspace{1.5mm}
\noindent
\textbf{Limitations and Future Work.} As our conclusions rely on observational data, potential confounders could limit the validity of our study. However, the robustness checks we conducted (content exact matching, sensitivity analysis) mitigate potential threats to our study's validity. 
Moreover, while our analysis centers on Reddit, where the subreddit structure offers an ideal context for our study, it's worth noting that prior research has stressed the significance of comprehending these mechanisms across various platforms~\cite{horta2021platform}. 
Future studies could extend this analysis to platforms lacking distinct communities, potentially yielding broader insight into the spread of fringe ideologies online.

\vspace{1.5mm}
\noindent
\textbf{Broader Impact,}
While our results suggest that curtailing fringe-interactions may effectively reduce the growth of fringe communities on mainstream platforms, it raises ethical concerns about potential censorship. We make two observations on that matter.
First, the ethical implications of limiting specific platform interactions deserve careful consideration. Balancing the need to mitigate harmful fringe communities with respect for free expression is a complex challenge, and platform stakeholders must weigh the potential harm of fringe communities against the principles of free speech.
Second, the narrative of "free speech" is a crucial mechanism that fringe platforms, such as Parler, use to lure users away from mainstream platforms. Limiting users' ability to post may result in migrations toward fringe platforms, putting users at risk of further radicalization.

\newpage

{ \small
\bibliography{refs.bib}
}

\section{Appendix}

\paragraph{Finetuning of BERT.}
We fine-tune  BERT \cite{devlin2018bert} with an additional linear layer to compute the propensity that a comment written by a user receives a fringe-interaction.
To do so, this model takes as input the content of the comment $c_{non-fringe}$ and the title of the corresponding Reddit submission. 
We, then, concatenate the [CLS] token with user and post-level covariates  to compute the propensity of a comment $c_{non-fringe}$ to receive a comment from a fringe user $u_{non-fringe}$

We fine-tune this model by following the recommendation of \citet{dodge2020fine}.
Specifically, we balance our training data including all comments $c_{non-fringe}$ written by a non-fringe user $u_{non-fringe}$ that received a fringe-interaction (treatment group) and a subsample of possible controls such as to build a balanced dataset for the training of BERT and the additional linear layer.
We then repeat the fine-tuning using five different random seeds for fifty different subsamples of the control group. 
We do not observe a statistical difference between the values of the loss function for the different subsamples.
We fine-tune an instance of this model for each of groups of interactions wit users of r/Incels, r/GenderCritical, and r/The\_Donald.
We trained each of these instances for 5 epochs in total. 
A Tesla T4 GPU for the fine-tuning of the model.
Finally, \Cref{tab:social_scores} shows a list of examples of comments written by treated and control users that that we matched using the model described above.

\paragraph{Robustness of Balanced Risk Set Matching} \label{app:robustness}

\begin{figure*}[t!]
    \centering
    \includegraphics[width=1.0\textwidth]{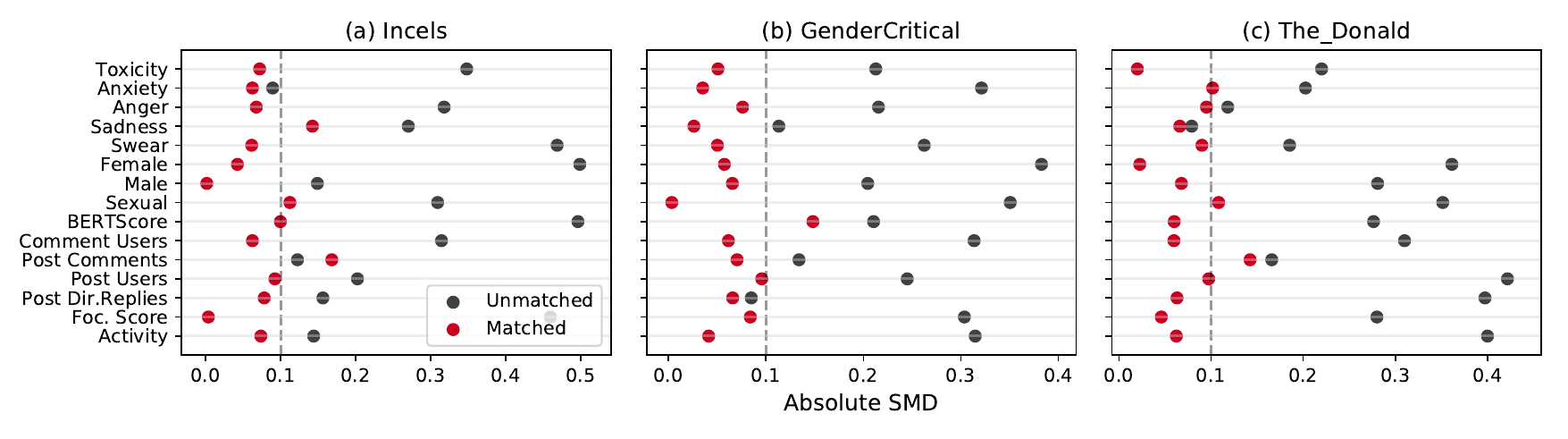}
    \caption{Absolute standardized mean differences (SMD) for fringe interactions before and after matching for all the covariates used for the balanced risk set matching. The SMDs have been computed for all three considered subreddits r/Incels, r/GenderCritical, and r/The Donald}
    \label{fig:matching}
\end{figure*}

We have evaluated the robustness of our results against two different matching procedure.
The first is described in \Cref{sec:methods:effect} which is based on propensity score matching and nearest neighbour algorithm.
The second approach we match treatment and control unitis directly in the covariates space. 
To do so, we compute the cosine similarity between the vectoral representations (i.e., [CLS]-token concatenated with post and user-level covariates) and match them using the nearest neighbor algorithm.
We do not find any statistically significant difference in our results using these two approaches. 
To show the quality of our matching, we show in \Cref{fig:matching} "love plots" for the absolute standardized mean differences of all post and user profile covariates described in \Cref{sec:methods:covariates}. 
To further check the average similarity of our matched comments at semantic level considering we measure the average BERTScore \cite{zhang2019bertscore} of the treatment comments with (i) the matched control comments and (ii) ten random samples of all possible control comments. 
Moreover, we also consider other linguistic traits like toxicity and LIWC dimensions to obtain a more interpretable representation of the similarity between matched comments.
We show the values of the absolute standardized mean errors (ASMD) before and after matching for all three subreddits we analyzed.

\paragraph{Subreddits Social Dimensions} \label{app:social_scores}
In \Cref{sec:rq3}, we stratify the effect of fringe-interactions across three "social dimensions" representing subreddits partisanship, age and gender.
We assign each subreddit to one of ten groups based on its social dimension score (deciles). 
\Cref{tab:social_scores} (bottom) provides a list of examples of subreddits in each of the decile for partisan, age and gender score computed accordingly to \citet{waller2019generalists}.

\paragraph{Collection of Communities Lingo.} \label{app:lingo}
To collect the lingo of r/Incels, r/GenderCritical, and r/The\_Donald specific community lingo we build a specialized web-crawler. 
We then collected from the offical websites of these three communities the list of the terms creating the glossary of these three communities.
We then count for each of the comments in our analysis the number of Incels, GenderCritical, and The\_Donald related terms. 


\begin{table*}[h]
    \centering
    \footnotesize
    \begin{tabularx}{\textwidth}{|X|X|}
        \hline
        \multicolumn{1}{|c|}{\textbf{Treatment}} & \multicolumn{1}{c|}{\textbf{Control}} \\
        \hline
        Maybe you are dressing or groomed so you look somewhat gay and why women aren't going out with you and men are hitting on you. Just a thought, no disrespect intended. & The whole point of being gay is that you are attracted to other men. Not men dressed as women or people born a woman and identify as a man and use a strap on. Sorry honey but maybe try Tinder \\
        \hline
        China's top politicians are mainly scientists, The USA's top politicians are mainly religious zealots and dumbfucks. Guess who invests more in climate change. & If anyone is going to speak of climate change, they must be armed with the true facts. Unfortunately, the major data points are very skewed. This was uncovered via Wikileaks. Follow the money. \\
        \hline
        Trump is the symptom of everything that is wrong with US politics while Hillary is the cause. Real talk that was worded beautifully, I wholeheartedly agree. & You're welcome, but please enlighten me on why Hillary should be president without mentioning Donald Trump. \\
        \hline
        If survival requires that we keep the environment to a certain level of stability, then we must. And what is your stance in the current pollution situation as well as big oil industry? & Yeah, because oil and gas companies are totally struggling for profits - notwithstanding the recent and temporary drop in oil prices. Also, you're aware that a focus on renewables would increase supply, which would decrease prices? It balances out and is certainly better for the environment. \\
        \hline
    \end{tabularx}  
    \caption{Examples of comments that received a fringe-interaction (treated) matched with comments that did not (control)}
    \label{tab:examples}
\begin{tabularx}{\textwidth}{|c|C|C|C|}
    \hline
    \textbf{Decile} & \textbf{Partisan} & \textbf{Age} & \textbf{Gender} \\
    \hline
    D1 & \textcolor[HTML]{053061}{democrats, Marijuana} &    \textcolor[HTML]{276419}{knifeparty, teenagers           } & \textcolor[HTML]{2d004b}{malelifestyle, fuckingmanly             }\\\hline
    D2 & \textcolor[HTML]{2166ac}{drugstore, TwoXSex} &      \textcolor[HTML]{4d9222}{skateboarding, Jokes            } & \textcolor[HTML]{542788}{SRSMen, Machinists                      }\\\hline
    D3 & \textcolor[HTML]{4393c3}{TVDetails, help} &         \textcolor[HTML]{7fbc41}{uncharted, studyAbroad          } & \textcolor[HTML]{8073ac}{DestructionPorn, Archery                } \\\hline
    D4 & \textcolor[HTML]{92c5de}{see, giantbomb} &          \textcolor[HTML]{b8e186}{ImaginaryMonsters, GirlTalk     } & \textcolor[HTML]{b2abd2}{porngifs, adorableporn                  } \\\hline
    D5 & \textcolor[HTML]{d1e5f0}{happygirls, MOMs} &        \textcolor[HTML]{e6f5d0}{Xsome, kickasstorrents          } & \textcolor[HTML]{d8daeb}{AdPorn, xxxcaptions                     } \\\hline
    D6 & \textcolor[HTML]{fddbc7}{Magic, porn} &             \textcolor[HTML]{fde0ef}{SugarBaby, KingstonOntario      } & \textcolor[HTML]{fee0b6}{weirdal, CryptoKitties                  } \\\hline
    D7 & \textcolor[HTML]{f4a582}{DebateCommunism, ideas} &  \textcolor[HTML]{f1b6da}{exmormon, MadMax                } & \textcolor[HTML]{fdb863}{PalaceClothing, WWE                     } \\\hline
    D8 & \textcolor[HTML]{d6604d}{FTC, BitcoinBeginners} &   \textcolor[HTML]{de77ae}{LifeProTips, vermont            } & \textcolor[HTML]{e08214}{gaymers, GirlsXBattle                   } \\\hline
    D9 & \textcolor[HTML]{b2182b}{Gunsforsale, russia} &     \textcolor[HTML]{c51b7d}{Plumbing, tuckedinkitties       } & \textcolor[HTML]{b35806}{DDLC, BattleCatsCheats                  } \\\hline
    D10 &\textcolor[HTML]{67001f}{Conservative, Military} &  \textcolor[HTML]{8e0152}{RedditForGrownups, MealPrepSunda} & \textcolor[HTML]{7f3b08}{againstmensrights, TheGirlSurvivalGuide } \\\hline
\end{tabularx}
    
    \caption{Examples of subreddits in each decile according to Partisan, Age and Gender Scores. Top rows are most left-leaning (\textcolor[HTML]{053061}{ \small \faSquare}), young (\textcolor[HTML]{276419}{\small \faSquare}), male (\textcolor[HTML]{542788}{\small \faSquare}) subreddits, respectively. Bottom rows are most right-leaning (\textcolor[HTML]{67001f}{\small \faSquare}), old (\textcolor[HTML]{8e0152}{ \small \faSquare}), and female (\textcolor[HTML]{7f3b08}{\small \faSquare}) subreddits, respectively. }
    \label{tab:social_scores}
\end{table*}


\end{document}